\documentstyle[letter,epsf]{ptptex}
\def\Peq{P^{\mbox{\scriptsize eq}}}
\def\lth{l_{\mbox{\scriptsize th}}}

\notypesetlogo  
\title{
A Geometrical Model for Stagnant Motion \\
in Hamiltonian Systems with Many Degrees of Freedom }

\author{
Yoshiyuki Y. {\sc Yamaguchi}
and Tetsuro {\sc Konishi}
}

\inst{
Department of Physics, Nagoya University, Nagoya 464-01
}

\abst{
  We introduce a model of Poincar\'e mappings which represents
  hierarchical structure of phase spaces for systems with many degrees 
  of freedom. 
  The model yields residence time distributions of a power type,
  and hence temporal correlation remains long.
  The power law behavior is enhanced as the system size increases.}

\begin{document}
\maketitle

\vspace*{0.25em}
\noindent {\it Introduction}
\hspace*{0.5em}
In many Hamiltonian systems, $1/f^{\nu} (0<\nu<2)$ power spectra and
long time tails have been observed, for instance, in area preserving
mappings,~\cite{karney83,chirikov84}
a water cluster,~\cite{baba97} and
a ferro-magnetic spin system.~\cite{yamaguchi97a} 
The $1/f^{\nu}$ spectra imply that relaxation to equilibrium is slow.
They are hence important phenomena of Hamiltonian systems with many
degrees of freedom.
We are interested in understanding the cause of $1/f^{\nu}$ spectra
from the structure of phase space and properties of motion.

To describe $1/f^{\nu}$ spectra,
Aizawa introduced a geometrical model for area preserving mappings,
which are models of Poincar\'e mappings for Hamiltonian systems with
two degrees of freedom.~\cite{aizawa84,aizawa89} 
This model assumes exact self-similar hierarchical structure of phase
spaces and produces stagnant motion, namely slow relaxation. 
We therefore understand that stagnant motion arises from self-similar
structure of phase space (often referred to ed as ``islands around
islands''~\cite{meiss86-islands}) and motion trapped to KAM tori or
Cantori. 
Meiss et al. have successfully proposed a similar
model.~\cite{meiss86-markovtree} 
A renormalization group approach,~\cite{escande85,hatori87}
which demonstrates similarity between scale transformations in phase
space and in time also supports the picture described above.
However, the models~\cite{aizawa84,meiss86-islands}
are based on the two-dimensionality of the phase space and 
cannot be directly applied to high dimensional systems.

For systems with many (more than two) degrees of freedom,
Aizawa et al. ~\cite{aizawa89} discussed the origin of the $1/f^\nu$
spectra based on the Nekhoroshev theorem.
Since the argument is based on the Nekhoroshev theorem,
the relation between stagnant motion and the hierarchical structure of 
phase space is not clear.

Moreover, the assumptions on which the models mentioned above are
based do not seem to hold for high-dimensional systems.
There exist many sorts of fixed points of Poincar\'e mappings
from the fully elliptic type to the fully hyperbolic type, 
and only the fully elliptic fixed points yield exact self-similarity
as area preserving mappings. 
Since the ratio of fixed points of the fully elliptic type decreases
as the system size becomes large, 
it seems impossible to assume exact self-similarity in the phase space 
structure for general high-dimensional Hamiltonian systems.
It is  believed that KAM tori rapidly disappear as the
systems size becomes large, and hence the volume of the region where
stagnant motion occurs also decreases. 
On the other hand, since
stagnant motion is frequently observed for Hamiltonian systems
with many degrees of freedom, we need to establish a model which 
yields stagnant motion in systems with many degrees of freedom
accordingly. 

In this paper, we propose a geometrical model which represents
hierarchical structure of phase spaces.
The model is an extension of Aizawa's model to many degrees of
freedom, and assumes that sticky zones exist around fixed points of
Poincar\'e mappings even if fixed points are not fully elliptic.
In other words, in systems with $N$ degrees of freedom, motion is
assumed to be trapped for a time around tori of fewer than $N$
dimensions also.

\vspace*{0.25em}
\noindent {\it Types of fixed points}
\hspace*{0.5em}
We consider Poincar\'e mappings $F$ and their fixed points instead of
Hamiltonian flows and their periodic orbits.
We set the number of degrees of freedom to $n$ for 
Poincar\'e mappings which have $2N-2$ dimensional Poincar\'e sections,
where $N$ is degrees of freedom of the Hamiltonian dynamics
(i.e. $n=N-1$).
Note that we construct a model based on fixed points of $F$ hereafter.
We can construct the model based on periodic points of $F$
with period-$k$ by using $F^k$ instead of $F$.

Local structure around fixed points is built from a combination of the 
three elementary types:
elliptic, hyperbolic and vortex types, for which the eigenvalues of
Jacobian of $F$ are $(e^{i\omega},e^{-i\omega})$, $(r,1/r)$ and
$(re^{i\omega},re^{-i\omega},e^{i\omega}/r,e^{-i\omega}/r)$
respectively, where both $\omega$ and $r$ are real.~ \cite{arnold-ap29}
The local structure is constructed as direct products of these three
types.

We assume that there is no vortex type structure for simplicity.
Generalization including the vortex type will be given in
Ref.~\citen{yamaguchi97b}.
Then local structure around a fixed point is constructed by elliptic
and hyperbolic types of structure, and there are $n+1$ varieties of
fixed points: direct products of $n-i$ elliptic type and $i$ hyperbolic
($i=0,1,\cdots,n$).
We define the index of a fixed point as $i$.
For instance, a fully elliptic fixed point is index-$0$ and a fully 
hyperbolic is index-$n$.

\begin{figure}[hbtp]
  \begin{center}
    \leavevmode
    \epsfxsize=14cm
    \epsffile{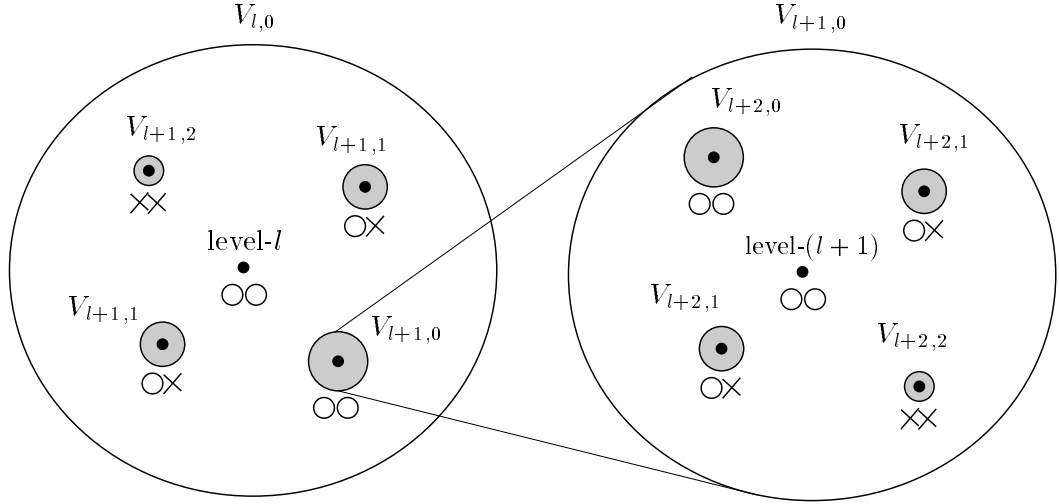}
    \caption{Schematic picture of the hierarchical structure of phase
      space. In this picture we assume the system size to be $n=2$.
      Black points are fixed points. 
      Circle and cross under the fixed points represent elliptic and
      hyperbolic elementary types, respectively.
      The double circle, for instance, implies that the structure
      around the fixed point is a direct product of two elliptic
      types, namely index-$0$.
      Let us focus on the left half of this figure.
      The fixed point at the center, whose index is $0$, belongs to 
      level-$l$, and the other four fixed points belong to
      level-$(l+1)$. 
      Shaded areas are sticky zones around fixed points of
      level-$(l+1)$.
      The sticky zone of the fixed point at the center is the inside
      of the biggest circle, which includes sticky zones of
      level-$(l+1)$ fixed points.
      Level-$(l+2)$ fixed points exist around level-$(l+1)$ fixed
      points, with a similar situation as on the right half,
      which is magnification around a fixed point with level-$(l+1)$
      and index-$0$.}
    \label{fig:hierarchy}
  \end{center}
\end{figure}

\vspace*{0.25em}
\noindent {\it Geometrical model and master equation}
\hspace*{0.5em}
Let us introduce a geometrical model with the following assumptions:\\
(G-1) Hierarchical structure is constructed by fixed points of
Poincar\'e mappings in phase spaces.\\
(G-2) Every sort of fixed points has a sticky zone around it,
even if it is not fully elliptic.\\
We calculate volumes of each level of hierarchy and derive a master
equation with some assumptions.
The number of the level is put in order by volume, and the base level
is level-$0$. 
We assume that the regions of level-$(l+1),(l+2),\cdots$ are in 
the region of level-$l$.
The schematic picture of this model is described in
Fig. \ref{fig:hierarchy}. 

Let us introduce notation for the quantities which will be used later:
\vspace*{-0.6em}
\begin{eqnarray}
N_{l,i}&:& \mbox{number of fixed points of level-}l,\mbox{ index-}i,
\nonumber \\ 
V_{l,i}&:& \mbox{volume of a sticky zone around a fixed point of
  level-}l,\mbox{ index-}i,
\nonumber \\
\hat{N}_{l}&:& \mbox{total number of fixed points of level-}l,
\nonumber \\ 
\hat{V}_{l}&:& \mbox{total volume of sticky zones of level-}l,
\nonumber \\ 
\rho_{l,i}&:& \mbox{number of sets of fixed points of level-}(l+1),
\nonumber \\ 
  &\hspace*{2em}& \mbox{which surround a fixed point of level-$l$
    index-$i$.} 
\nonumber
\end{eqnarray}\\

\vspace*{-2em}
\noindent Here we have assumed:\\
(G-3) Fixed points have the same volume $V_{l,i}$ if their level and
index are the same.\\
Note $\hat{V}_{l} > \hat{V}_{l+1}$ for all $l$ because we require
that $\hat{V}_{l}$ includes $\hat{V}_{l+1},\hat{V}_{l+2},\cdots$.
The meaning of $\rho_{l,i}$ is clarified in the following.

\vspace*{0.3em}
(Fact): A fixed point of level-$l$ and index-$i$ is surrounded by
$\rho_{l,i} ({}_{n-i} C_{j})$ fixed points of level-$(l+1)$ and
index-$(i+j)$, where $j=0,1,\cdots,n-i$, and $\rho_{l,i}$ is a
positive integer. 
\vspace*{0.3em}

This fact is an extension of the Poincar\'e-Birkhoff
theorem~\cite{lichtenberg92} for many degrees of
freedom.~\cite{arnold-ap33}
Using this fact, we write the recursion formula for $N_{l,i}$ as
\begin{equation}
  N_{l+1,i} = \sum^i_{k=0} \rho_{l,k} N_{l,k} ({}_{n-k} C_{i-k}).
  \label{number}
\end{equation}

The total number $\hat{N}_{l}$ and volume $\hat{V}_{l}$ of the
level-$l$ sticky zones are
\begin{equation}
  \hat{N}_{l} = \sum^{n}_{i=0} N_{l,i}, \qquad
  \hat{V}_{l} = \sum^{n}_{i=0} N_{l,i} V_{l,i}.
  \label{total}  
\end{equation}

To observe motion among levels we introduce a master equation with
the following three assumptions:\\
(M-1) Systems are ergodic. \\
(M-2) Transitions from level-$l$ are limited to level-$(l-1)$,
  $l$ and $(l+1)$. \\
(M-3) Transitions among levels are Markovian. \\ 
From (M-1), the probability being level-$l$ in equilibrium,
$\Peq_{l}$, is proportional to volume $\hat{V}_{l}$:
\begin{equation}
  \Peq_{l} \propto \hat{V}_{l}.
  \label{equilibrium}
\end{equation}
We assume detailed balance to fix the transition probability
$w_{m,l} \equiv w\{l\to m\}$. Then
\begin{equation}
  w_{l,m}\Peq_{m} = w_{m,l}\Peq_{l}.
  \label{detailed-balance}
\end{equation}
From (M-2) and Eqs. (\ref{equilibrium}) and
(\ref{detailed-balance}), transition probabilities are written as
\begin{equation}
  w_{m,l} =  \left\{
    \begin{array}{ll}
      c \hat{V}_{l-1},           & (m=l-1) \\
      1- w_{l-1,l} - w_{l+1,l},  & (m=l) \\
      c \hat{V}_{l+1},           & (m=l+1) \\
      0.                         & (\mbox{otherwise})
    \end{array}
    \right.
  \label{trans-prob}
\end{equation}
The factor $c$ is independent of the level, and we set
$c=1/(2(\hat{V}_0+\hat{V}_2))$.
By using these transition probabilities, the master equation is
written as
\begin{equation}
  P_{l}(t+1) = \sum_{m} w_{l,m} P_{m}(t),
  \label{Master-eq}
\end{equation}
where $P_{l}(t)$ is the probability of being on level-$l$ at step
$t$.

\vspace*{0.25em}
\noindent {\it Results of numerical calculations}
\hspace*{0.5em}
We numerically calculated residence time distributions based on our
model Eq. (\ref{Master-eq}) and examined if it obeys a power law. 
The residence time distribution $R(t)$ is the probability that motion
extends to levels shallower than level-$\lth$ for the first time with
initial level being $\lth$ at $t=0$.
We obtain $R(t)$ from the following transition probabilities and initial
condition: 
\begin{equation}
  w_{m,\lth-1} = 0 \quad (m=\lth-1,\lth,\lth+1),
  \qquad P_{l}(t=0) = \delta_{l,\lth}.
\end{equation}
Here $\delta_{l,\lth}$ is Kronecker's delta. Then we have
\begin{equation}
  R(t) = P_{\lth-1}(t).
\end{equation}
If $R(t)$ is of a power type rather than an exponential, motion among
levels is stagnant. 

Parameters which must be given are assumed as follows:
\[
  \rho_{l,i} = 2, \qquad
  V_{l,i} = b^{-(l+i)}, \qquad
  \lth = 1,
\]
\[
  N_{0,0} = 1 \quad \mbox{and}\quad N_{0,i} = 0.\quad(i>0)
\]
Sticky zones become small as hyperbolic components increase,
and this index dependence of $V_{l,i}$ is an essential point of our
model. 
We assumed that the volume of a sticky zone $V_{l,i}$ decays as
$b^{-i}$ with respect to the index-$i$ of the fixed point,
since local structure near fixed points is constructed by direct
products.
Other forms of $V_{l,i}$, for instance
$V_{l,i}=b^{-l}\tilde{b}^{-i}$ and $V_{l,i}=\exp[-(l+i^{\alpha})\ln b]$,
give similar results to $V_{l,i} = b^{-(l+i)}$ with appropriate
values of parameters.
The form of $V_{l,i}$ determines which index is dominant in
$\hat{V}_l$.

\begin{figure}[hbtp]
  \begin{center}
    \leavevmode
    \epsfxsize=7cm
    \epsffile{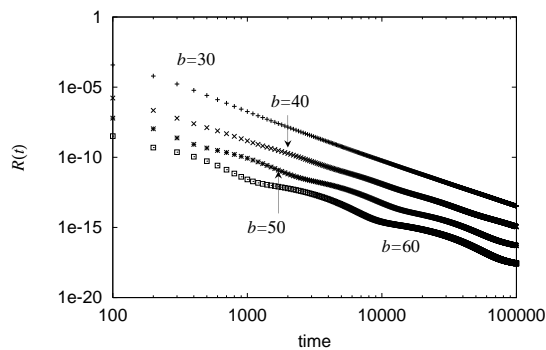}
    \caption{Residence time distributions for various values of the
      scale factor $b$. Here $n=80$. The magnitude of the longitudinal 
      axis is multiplied by $10^{-1}, 10^{-2}$ and $10^{-3}$ for
      $b=40,\ 50$ and $60$, respectively. } 
    \label{fig:n=80}
  \end{center}
\end{figure}

The residence time distribution $R(t)$ is shown in Fig. \ref{fig:n=80} 
for $n=80$ and various values of $b$.
When $b$ is small ($b=30,40$) $R(t)$ is of a power type. Namely, we
have 
\begin{equation}
  R(t) \sim t^{-\beta},
\end{equation}
where $\beta$ is $3.4$ and $3.1$ for $b=30$ and $b=40$, respectively.
Since $R(t)$ is of a power type, stagnant motion occurs among levels.
The stagnant motion is also observed when $n=1$, and hence our model 
Eq. (\ref{Master-eq}) is consistent with models for area preserving
mappings mentioned in the Introduction.
The values of $\beta$ are also consistent with those obtained in
area preserving mappings which are models of physical systems and give 
$1.5\leq\beta\leq 3$. \cite{karney83,chirikov84,irie54}
As far as we know, $\beta$ has not been calculated in
Hamiltonian systems with many degrees of freedom.
Appearance of oscillating behavior for $b=50$ and $b=60$ is caused
by the weakness of effects of the hierarchy, which become weaker as
$b$ increases, because large $b$ implies that level-$(l+1)$ is small
compared with level-$l$. 

\begin{figure}[hbtp]
  \begin{center}
    \leavevmode
    \epsfxsize=7cm
    \epsffile{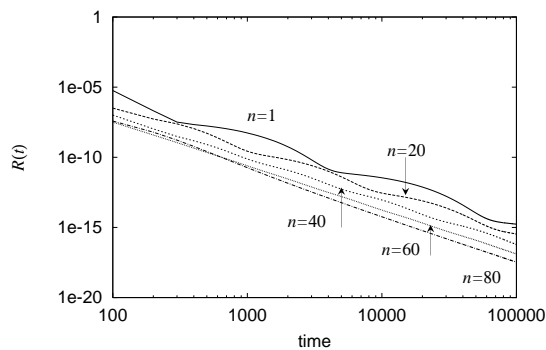}
    \caption{Dependence on the number of degrees of freedom of
      residence time distributions with fixed $b=30$. 
      The magnitude of the longitudinal axis is multiplied by
      $10^{-1},\ 10^{-2},\ 10^{-3}$ and $10^{-4}$ for $n=20,\ 40,\ 60$ 
      and $80$, respectively. The distributions are closer to a
      power type as $n$ increases.}
    \label{fig:n-dep}
  \end{center}
\end{figure}

We display the residence time distribution $R(t)$ for various system
size $n$ in Fig. \ref{fig:n-dep}, where power law behavior of $R(t)$
is clearly seen.
Oscillations found for small $n$ gradually decay as $n$ 
becomes large, and the distributions are close to a power type as $n$
increases. 
This is an effect of the many degrees of freedom and indicates that
fine 
tuning of parameters is not necessary to observe $1/f^{\nu}$ spectra
in systems with many degrees of freedom.

\vspace*{0.25em}
\noindent {\it Summary}
\hspace*{0.5em}
To understand $1/f^{\nu}$ spectra and long time tails in Hamiltonian
systems with many degrees of freedom,
we proposed a geometrical model of phase space, which is an extension
of Aizawa's model.
We assumed that sticky zones exist around fixed points of Poincar\`e
mappings even if the fixed points are not fully elliptic,
and accordingly, exact self-similarity of phase space is not
introduced. 
We derived a master equation from our model, and found that residence
time distributions are of a power type. 
That is, stagnant motion among levels occurs although phase space does 
not possess exact self-similarity.
The power law behavior becomes clearer as the system size increases.

\vspace*{0.5em}
We express our thanks to members of R-lab. of Nagoya University for
useful discussions.


\end{document}